\documentclass[epsfig,12pt,onecolumn]{article}
\usepackage{amsfonts}
\usepackage{amssymb}
\usepackage{amsmath}
\usepackage{multicol}
\usepackage{graphicx}
\usepackage{float}
\usepackage{caption}

\makeatletter \@addtoreset{equation}{section}

\def \be{\begin{equation}}
\def \ee{\end{equation}}
\def \bea{\begin{eqnarray}}
\def \eea{\end{eqnarray}}

\newcommand{\nc}{\newcommand}
\nc{\al}{\alpha} \nc{\bib}{\bibitem} \nc{\la}{\lambda}
\nc{\C}{\mbox{\hspace{1.24mm}\rule{0.2mm}{2.5mm}\hspace{-2.7mm} C}}
\nc{\R}{\mbox{\hspace{.04mm}\rule{0.2mm}{2.8mm}\hspace{-1.5mm} R}}

\begin{document}

\title{ Preheating and Reheating after Standard Inflation }
\author{K. El Bourakadi$^{1}$\thanks{%
k.elbourakadi@yahoo.com } \\
%EndAName
$^{1}${\small Quantum Physics and Applications Team, LPMC,}\\
{\small \ Ben M'sik Faculty of Sciences, Hassan II University, Casablanca,
Morocco}}
\maketitle

\begin{abstract}
We study the two-phase scenario following inflation, where the initial step
is preheating, accompanied by a step of perturbative reheating at which
inflaton field decays transferring all of its energy to create relativistic
particles, the interaction of these particles will evolve towards a state of
thermal equilibrium with a temperature $T_{re} $ called the reheating
temperature. It is observed that the scenario of reheating normally predicts
the maximum reheating temperature that corresponds to an almost
instantaneous transition from inflation to the radiation domination era.
This will naturally lead to a nonperturbative preheating. In this framework,
we propose constraints on the reheating duration parameters expressed in
terms of the cosmic microwave background (CMB) inflationary scalar spectral
index. In this work we study the compatibility of polynomial and Higgs
models of inflation with the observational data obtained from Planck 2018.
\end{abstract}

\section{Introduction}

At the end of inflation, the universe enters a new stage, at which, the
creation of elementary particles and its decay occurs. Note that, in the
first phase, the creation of particles occurs due to parametric resonance
called preheating, and in the last phase, the particles decay and thermalize
with a final temperature called reheating temperature. The inflaton field $%
\phi $ starts oscillating near the minimum of its potential, with a decrease
of the amplitude of the oscillations over time because the inflaton's energy
is transferred to other fields and relativistic particles. This is the
mechanism of the production of elementary particles following inflation
which we call reheating, the produced particles interact with each other
until they reach a temperature of thermal equilibrium $T_{re}$.

At the stage of preheating\cite{Kofman}, the field decay into bosons and
because of the Pauli exclusion principle fermions production occurs only at
the late stage of reheating, knowing that reheating at the parametric
resonance stage is never complete, at the start the resonance was broad but
eventually, it became narrow \cite{2} then the reheating ends at the time of
the thermalization of produced particles. In \cite{instant1,instant2} models
of inflation considering instant reheating have been tested. We, therefore,
will take into account the maximum reheating temperature that leads to the\
lower values of $N_{re}$ wich favor a much higher e-folds number of
preheating, and prove that it's parameterized by a post-inflationary e-folds 
$N_{pre}.$ The equation of state (eos) $\omega $ is an important parameter
that can be used in the study of the evolution of the universe, the (eos)
takes different values in each era of its history, for example, $\omega \geq
-1/3$ is needed to end inflation\cite{Dai,Asadi} but during the reheating
the (eos) increase from $0$ to $1/3$, where $\omega =1/3$ corresponds to a
universe of the radiation domination era\cite{Cook}.

\section{Parametric resonance}

The reheating process occurs when the inflaton energy density is transferred
to the energy density of other fields, The decay rate of the inflaton
oscillations parametrized with $\Gamma $\cite{Kofman}, is added as a
friction term to the motion equation given by \ eq.(1)\ \ which proves to be
very useful in the study of reheating after inflation :

\begin{equation}
\ddot{\phi}+3H(t)\dot{\phi}+\Gamma \dot{\phi}+m^{2}\phi =0.  \label{aq1}
\end{equation}%
The solution of this\ equation from \cite{Schmitz} suggested the concept
that one can explain the effects of reheating using the total decay rate
given by the formula $\Gamma =\Gamma (\phi \rightarrow \chi \chi )+\Gamma
(\phi \rightarrow \overline{\psi }\psi )$, which describe the energy
transferring to the new particles. 
\begin{equation}
\phi (t)\approx \Phi (t)sin(m_{\phi }t),
\end{equation}
$\phi (t)$ is the solution of the first equation, and $\Phi (t)$ is a
decreasing amplitude that is in the form : 
\begin{equation}
\Phi (t)=\Phi _{0}exp\left[ -\dfrac{1}{2}(3H+\Gamma )t\right] .
\end{equation}

Preheating is the first stage of the reheating era, which is characterized
by the appearance of $\phi $ and $\chi $ particles\cite{Kofman}, which
cannot be studied using perturbative theory\cite{2}. Oscillations in the
scalar field can decay into light $\chi $-particles, these particles\ are
taken into account because of the interaction $\dfrac{1}{2}g^{2}\phi
^{2}\chi ^{2}$ which is given by the potential of the form 
\begin{equation}
V(\phi ,\chi )=\dfrac{1}{2}m_{\phi }^{2}\phi ^{2}+\dfrac{1}{2}m_{\chi
}^{2}\chi ^{2}+\dfrac{1}{2}g^{2}\phi ^{2}\chi ^{2},
\end{equation}%
the time evolution of the quantum fluctuation of the field $\chi $ is given
by 
\begin{equation}
\ddot{\chi}+3H\dot{\chi}-\dfrac{1}{a^{2}}\nabla _{x}^{2}\chi +V^{^{\prime
}}(\phi ,\chi )=0,
\end{equation}%
deriving the potential with respect to $\chi $ and proceeding into Fourier
space : 
\begin{equation}
\ddot{\chi}_{k}+3H\dot{\chi _{k}}+\left( \dfrac{k^{2}}{a^{2}}+m_{\chi
}^{2}+g^{2}\phi ^{2}(t)\right) \chi _{k}=0,
\end{equation}%
according to\cite{Kofman} the above equation describes $\chi $-particles
production by the excitation of the field, these particles have momentum $%
k/a\gg m_{\phi }$, knowing that the inflaton field only interacts with a
light scalar field$\ \chi $ under the condition ($m_{\chi }\ll m_{\phi }$),
taking in consideration the constant expansion of the universe $H\approx 0$,
and using the solution in eq.19 with the condition that $\Phi $ is varying
slowly compared to the oscillation of $\chi $-field\cite{Schmitz}, the eq.23
is reduced to : 
\begin{equation}
\ddot{\chi}_{k}+\left( \dfrac{k^{2}}{a^{2}}+g^{2}\Phi ^{2}sin^{2}(m_{\phi
}t)\right) \chi _{k}=0,
\end{equation}%
using $z=mt$ and $sin^{2}(z)=1/2(1-cos(2z))$, we get the well known Mathieu
equation : 
\begin{equation}
\chi _{k}^{\prime \prime }+(A_{k}-2qcos(2z))\chi _{k}=0,
\end{equation}%
with $A_{k}=k^{2}/m^{2}+2q$, $q=g^{2}\Phi ^{2}/4m^{2}$ and prime denote
differentiation with respect to $z$. this equation describes an oscillator
with a periodically changing frequency $\omega ^{2}(t)=k^{2}+g^{2}\Phi
^{2}sin(mt)$. As already mentioned, in the parametric resonance regime, the
bosons were created due to the broad parametric resonance then the resonance
became narrow. The broad resonance is known as the case were the parameter $%
q\gg 1$, and for $g\phi <m$ we have the narrow resonance with $q\ll 1$. An
important feature of the solution of Mathieu equation is the existence of an
exponential instability $\chi _{k}\propto exp(\mu _{k}z)$, this instability
corresponds to an exponential growth of occupation number of quantum
fluctuation $n_{k}(t)\propto exp(2\mu _{k}z)$\cite{2}.

\section{Reheating duration}

\bigskip Information about reheating can be extracted considering the
history of the Universe between inflation at which observed CMB modes
crossed beyond the Hubble radius and the present time. Knowing that
deferents eras occurred throughout this length of time that are parametrized
by several e-folds numbers, 
\begin{equation}
\dfrac{k}{a_{0}H_{0}}=\dfrac{a_{k}}{a_{end}}\dfrac{a_{end}}{a_{re}}\dfrac{%
a_{re}}{a_{eq}}\dfrac{a_{eq}H_{eq}}{a_{0}H_{0}}\dfrac{H_{k}}{H_{eq}}
\end{equation}%
\newline
\begin{equation}
ln\frac{k}{a_{0}H_{0}}=-N_{k}-N_{re}-N_{RD}+ln\frac{a_{eq}H_{eq}}{a_{0}H_{0}}%
+ln\frac{H_{0}}{a_{eq}}
\end{equation}%
here $k$ refers to the pivot scale for a specific experiment\cite{Cook}, and 
$N_{k}$ is the e-folds of the inflation era, $N_{re}$ and $N_{RD}$
respectively correspond to the reheating and radiation-domination era
durations.

In addition to its thermalization temperature $T_{re}$ and equation of state 
$\omega _{re}$, reheating is also characterized by the number of e-folds $%
N_{re}=ln(a_{re}/a_{end})$ occurring between the time inflation ends, and
the beginning of the radiation-dominated era, which we called the duration
of reheating. Using $\rho \propto a^{-3(1+\omega _{re})}$, the reheating era
can be described as :\newline
\begin{equation}
\dfrac{\rho _{end}}{\rho _{re}}=\left( \dfrac{a_{end}}{a_{re}}\right)
^{-3(1+\omega _{re})}
\end{equation}%
where $\rho _{end}$ and $a_{end}$ corresponds to the end of inflation and $%
\rho _{re}$ and $a_{re}$ corresponds to the end of reheating. We can
re-write this using e-folds equation as\ 
\begin{equation}
N_{re}=\dfrac{1}{3(1+\omega _{re})}ln\left( \dfrac{\rho _{end}}{\rho _{re}}%
\right) ,
\end{equation}%
and replace the energy densities by their expressions\cite{Cook}, $\rho
_{end}=\dfrac{3}{2}V_{end}$ and $\rho _{re}=\dfrac{\pi ^{2}}{30}%
g_{re}T_{re}^{4}$ in the previous equation to have : 
\begin{equation}
N_{re}=\dfrac{1}{3(1+\omega _{re})}ln\left( \dfrac{3}{2}\dfrac{30V_{end}}{%
\pi ^{2}g_{re}T_{re}^{4}}\right) ,
\end{equation}%
the conservation of the entropy between the end of reheating and the actual
time can lead us to the assumption that reheating temperature and today
temperature are related as\cite{Dai} : 
\begin{equation}
T_{re}=T_{0}\left( \dfrac{a_{0}}{a_{eq}}\right) e^{N_{RD}}\left( \dfrac{43}{%
11g_{re}}\right) ^{1/3},
\end{equation}%
as a starting point, we will include the next equation%
\begin{equation}
\dfrac{a_{0}}{a_{eq}}=\dfrac{a_{0}}{a_{k}}\dfrac{a_{k}}{a_{end}}\dfrac{%
a_{end}}{a_{re}}\dfrac{a_{re}}{a_{eq}},
\end{equation}%
the expression of the previous equation in terms of e-folds number will be
given by : 
\begin{equation}
\dfrac{a_{0}}{a_{eq}}=\dfrac{a_{0}H_{k}}{k}e^{-N_{k}}e^{-N_{re}}e^{-N_{RD}},
\end{equation}%
the expression of the temperature in the eq. (3.6) became : 
\begin{equation}
T_{re}=\left( \dfrac{T_{0}a_{0}}{k}\right) H_{k}e^{-N_{k}}e^{-N_{re}}\left( 
\dfrac{43}{11g_{re}}\right) ^{1/3},
\end{equation}%
replacing the previous equation in eq. (15) gives : 
\begin{equation}
N_{re}=\dfrac{4}{1-3\omega _{re}}\left[ \dfrac{1}{3}ln\left( \dfrac{11g_{re}%
}{43}\right) -ln\left( \dfrac{k}{a_{0}T_{0}}\right) -ln\left( \dfrac{%
V_{end}^{1/4}}{H_{k}}\right) -N_{k}\right] .
\end{equation}%
\newline
Considering $g_{re}\approx 100$ and the pivot scale $0.05Mpc$ \cite{Cook},
the numerical application gives : 
\begin{equation}
N_{re}=\dfrac{4}{1-3\omega _{re}}\left[ 6.61-ln\left( \dfrac{V_{end}^{1/4}}{%
H_{k}}\right) -N_{k}\right] .
\end{equation}

Now it is possible to calculate reheating duration as a function of the tree
parameters $V_{end}$, $H_{k}$ and $N_{k}$ under the condition that $N_{re}$
is known. according to\cite{Sakhi}, the reheating ended at a temperature in
the order of $10^{14}Gev$, if we used this result and applied it in eq.(15)
we can find the reheating duration. Relating these parameters\ with the
scalar spectral index $n_{s}$, and the choice of a specific inflation model
with reheating (eos) $\omega _{re}$, will lead us to constrain reheating
parameters in the standard inflation.

\section{Reheating constraints in standard inflation}

\subsection{Standard infation}

The main purpose of the theory of inflation is to solve the problems related
to the Big-bang notably the horizon, flatness, and monopole problems.
According to this theory, a scalar field called inflaton was responsible for
an explosive phase that made the global geometry of the universe flat and
made all regions of the universe connected causally. The evolution of the
scalar field $\phi $ is described with the Klein-Gordon equation $\ddot{\phi}%
+3H\dot{\phi}+V^{\prime }(\phi )=0$. For a successful inflation, the
slow-roll parameters given by $\epsilon =M_{p}^{2}V^{\prime 2}/2V^{2}$ and $%
\eta =M_{p}^{2}V^{\prime \prime }/V$ must obey the conditions $\epsilon <<1$
and $\eta <<1$.\newline

Since the tree parameters $N_{k}$, $V_{end}$ and $H_{k}$ can be calculated
as a function of the potential $V$ that can be related also to the slow-roll
parameters, these parameters can be tested experimentally if we relate the
slow-roll parameters to the scalar spectral index as described in the next
equation: 
\begin{equation}
n_{s}-1=-6\epsilon +2\eta
\end{equation}%
Once the form of $V(\phi )$ is specified for a given model, we can compute $%
N_{re}$ as a function of inflationary parameter $n_{s}$ for $\omega _{re}\in
\lbrack -1/3,0]$

\section{Polynomial potentials}

Consider a polynomial potential 
\begin{equation}
V=\dfrac{1}{2}m^{4-n}\phi ^{n}.
\end{equation}%
We first need to calculate the parameters $N_{k}$, $H_{k}$, and $V_{end}$.
The number of e-folds between the time the pivot scale exited the Hubble
radius and the end of inflation is given by

\begin{equation}
N_{k}\simeq \dfrac{1}{M_{p}^{2}}\int_{\phi _{end}}^{\phi _{k}}\dfrac{V}{%
V^{\prime }}d\phi ,
\end{equation}%
applying the previous equation to the polynomial potential :

\begin{equation}
N_{k}=\dfrac{1}{4nM_{p}^{2}}(\phi _{k}^{2}-\phi _{end}^{2}),
\end{equation}%
since $\phi _{end}$ is very small compared to the field at the beginning and
during the inflation era, $\phi _{end}<<\phi _{k}$ it is appropriate to
approximate 
\begin{equation}
N_{k}\approx \dfrac{1}{4nM_{p}^{2}}\phi _{k}^{2},
\end{equation}%
we must write $N_{k}$ as a function of $n_{s}$ using the slow-roll
parameters mentioned in the previous section and eq. (22) to obtain : 
\begin{equation}
N_{k}=\dfrac{n+2}{2(1-n_{s})}.
\end{equation}%
Knowing that the Friedmann equation is written as :

\begin{equation}
H_{k}^{2}=\dfrac{1}{3M_{p}^{2}}V_{k},
\end{equation}

\begin{equation}
r_{k}=\dfrac{2H_{k}^{2}}{\pi ^{2}M_{p}^{2}A_{s}},
\end{equation}%
then using $r=16\epsilon _{k}$ gives 
\begin{equation}
H_{k}=\pi M_{p}\sqrt{\dfrac{4nA_{s}}{n+2}(1-n_{s})},
\end{equation}%
and finally we can find $V_{end}$ \cite{Dai} using 
\begin{equation}
V_{k}=V_{end}\left( \dfrac{\phi _{k}}{\phi _{end}}\right) ^{n}
\end{equation}%
to compute $V_{end}$ in terms of $n_{s}$ and $A_{s}$ \ 
\begin{equation}
V_{end}=\frac{12}{\sqrt{2}}\pi ^{2}M_{p}^{4}A_{s}\left( \dfrac{n(1-n_{s})}{%
(n+2)}\right) ^{n+2},
\end{equation}%
using $n_{s}=0.9649\pm 0.0042$ and Planck's central value $\ln
(10^{10}A_{s})=3.044$, we can plot the variations of $N_{re}$ as a function
of $n_{s}$ for four different values of $\omega _{re}$ (see Figs. 1 and 2). 
\begin{figure}[]
\centering
\includegraphics[width=15cm]{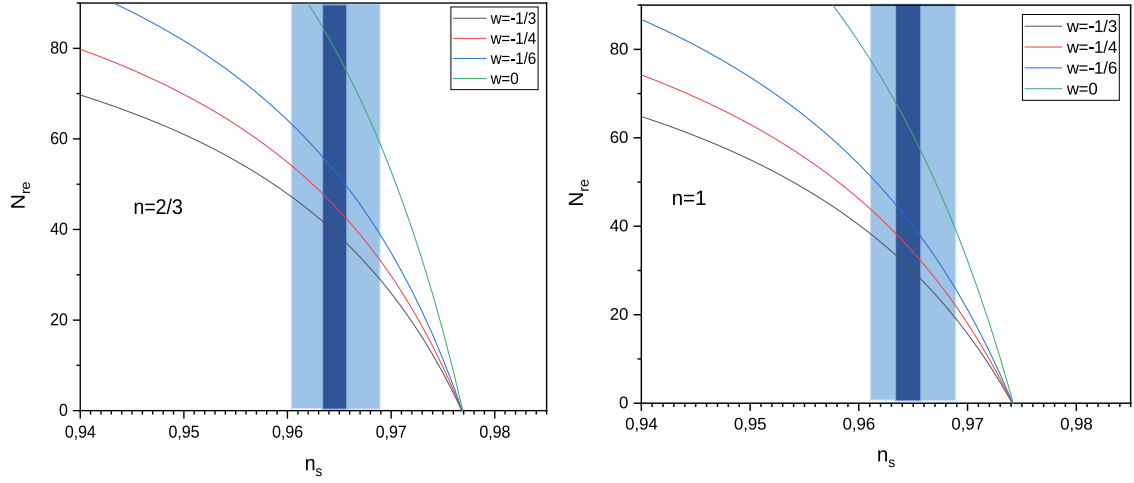}
\caption{Variation of $N_{re}$ as function of $n_{s}$ for $n=2/3$ and $n=1$
for differentes values of $\protect\omega _{re}$.}
\label{graph1}
\end{figure}
\begin{figure}[]
\centering
\includegraphics[width=15cm]{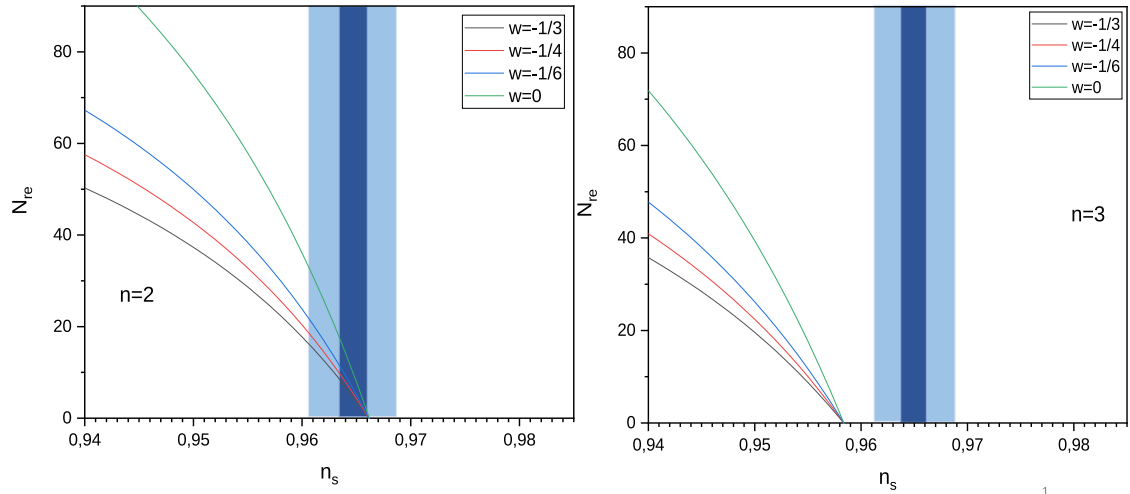}
\caption{Variation of $N_{re}$ as function of $n_{s}$ for $n=2$ and $n=3$
for differentes values of $\protect\omega _{re}$.}
\label{graph2}
\end{figure}

In Fig. 1 and 2, we apply the previous results to compute $N_{re}$ as
functions of $n_{s}$. choosing the polynomial potential we study the
reheating duration for several values of $n=2/3,1,2,3$. Moreover, we focus
on reheating equation-of-state parameters in the interval $[-1/3,0]$. We
plot $N_{re}$ as a function of $n_{s}$ for different values of the
preheating equation-of-state parameters, in each case: $\omega _{re}=-1/3$
(black line), $\omega _{re}=-1/4$ (red line), $\omega _{re}=-1/6$ (blue
line), and $\omega _{re}=0$ (green line). \newline
We observe that for all four cases, all the lines with different values of $%
\omega _{re}$ converge to a central value where the preheating is
instantaneous $N_{re}\rightarrow 0$, when we compared the results of the
four cases we find that the chaotic model with $n=2$ is the most compatible
with Planck's $1\sigma $ bounds on $n_{s}$, but the case $n=3$ for all $%
\omega _{re}$ is difficult to reconcile in $1\sigma $ on $n_{s}$. As for the
chaotic potential, the (eos) value$\ \omega _{re} \simeq 0\ $is needed for
an efficient reheating because it enables a much prolonged e-folds number of
reheating.

\section{\protect\bigskip Higgs potential}

\bigskip The Higgs inflation is based on the idea that the Higgs field is
the inflaton field by adding a non-minimal coupling to gravity, the Higgs
field potential is given by : 
\begin{equation}
V=M_{p}^{4}\left( 1-e^{-\phi \sqrt{2/3}/M_{p}}\right) ^{2},
\end{equation}%
using this potential, one can calculate the number of e-foldings during
inflation as :

\begin{center}
\begin{equation}
N_{k}=\dfrac{1}{M_{p}}\int_{\phi _{end}}^{\phi _{k}}\dfrac{V}{V^{\prime }}%
d\phi =\dfrac{1}{2M_{p}^{2}}\sqrt{\dfrac{3}{2}}\left[ M_{p}\sqrt{\dfrac{3}{2}%
}e^{\phi \sqrt{2/3}/M_{p}}-\phi \right] _{\phi _{end}}^{\phi _{k}}.
\end{equation}
\end{center}

Knowing that the field obeys the condition $\phi _{k}\gg \phi _{end}$, and
using the condition $M_{p}\sqrt{\dfrac{3}{2}}e^{\phi \sqrt{2/3}/M_{p}}\gg
\phi _{k}$. the number of e-folds is simplified to 
\begin{equation}
N_{k}=\dfrac{3}{4}e^{\phi \sqrt{2/3}/M_{p}},
\end{equation}%
which implies 
\begin{equation}
\phi _{k}=\sqrt{\dfrac{3}{2}}M_{p}ln\left( \dfrac{4}{3}N_{k}\right) ,
\end{equation}%
using eq.(34) and the slow-roll parameters that are expressed by : 
\begin{equation}
\epsilon _{k}\simeq \dfrac{3}{4N_{k}^{2}},\eta _{k}=-\dfrac{1}{N_{k}}
\end{equation}%
one must find the inflation e-folds which is given by 
\begin{equation}
N_{k}=\dfrac{2}{1-n_{s}},
\end{equation}%
combining the expressions above, one derives $H_{k}$ and $V_{end}$ as
function of $n_{s}$ and $A_{s}$ in the form :%
\begin{equation}
H_{k}=\pi M_{p}\sqrt{\dfrac{3}{2}A_{s}}(1-n_{s})
\end{equation}
and 
\begin{equation}
V_{end}=\dfrac{9}{2}\pi ^{2}M_{p}^{4}A_{s}(1-n_{s})^{2}\dfrac{\left( \dfrac{1%
}{\sqrt{3/2}+1}\right) ^{2}}{\left( 1-3/8)(1-n_{s})\right) ^{2}}
\end{equation}%
All that remains now is to calculate the reheating duration for the case of
Higgs inflation, taking in consideration the central values $n_{s}=0.9649\pm
0.0042$ and $\ln (10^{10}A_{s})=3.044$ according to Plank's 2018 data. 
\begin{figure}[]
\centering
\includegraphics[width=14cm]{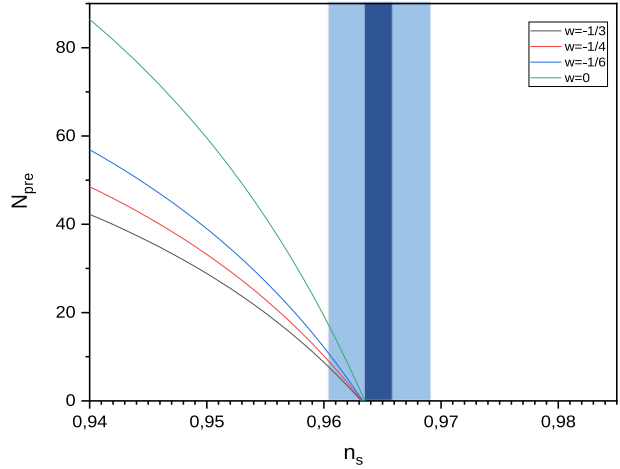}
\caption{Variation of $N_{re}$ as function of $n_{s}$ in the case of Higgs
inflation for $n=2/3$ and $n=1$ for differentes values of $\protect\omega %
_{re}$.}
\label{graph2}
\end{figure}

The result in Figure 3 is a computation of $N_{re}$ as functions of $n_{s}$.
Choosing the Higgs potential we study the reheating duration and focus on
the same interval of (eos) parameters, We plot $N_{re}$ as a function of $%
n_{s}$. \newline

The point where all the lines converge is considered to be the point of
instantaneous reheating which is defined as the limit $N_{re}\rightarrow 0. $
We observe that for all different values of $\omega _{re}$ this model is
compatible with the Planck $1\sigma $ bounds on $n_{s}$ because all the
lines in this Figure are shifted toward the central value of $n_{s}$.
Comparing this case with the chaotic inflation, the Higgs model (eos)
parameter $\omega _{pre} \simeq 0\ $is also required for an Efficient
reheating since it allows a much prolonged duration of this stage.

\section{Conclusion}

In this work, we have calculated the reheating duration in the standard
inflation for various cosmological parameters. We studied the preheating
phase for the polynomial potential with the form $V\propto m^{4-n}\phi ^{n}$
and the Higgs inflation. We showed that for the polynomial potential, the
chaotic model with $n=2$ provides the best fit results with recent
observational constraints because it's compatible with Planck's results,\
but the case $n=3$ shifted away from observation. The Higgs model shows good
compatibility with Planck's bounds of $n_{s}$. We finally concluded that for
both potentials effecient reheating demand that $\omega _{re}\simeq 0$ which
favors a much more post-inflationary e-folds of expansion for the reheating
phase.

\end{document}